\documentclass[conference]{IEEEtran}
\usepackage{booktabs}
\usepackage{bm}
\usepackage{graphicx}
\usepackage[cmex10]{amsmath}
\usepackage{cite}

\newtheorem{proposition}{Proposition}
\newtheorem{corollary}{Corollary}

\begin{document}

\title{Constellation Mapping for Physical-Layer Network Coding with M-QAM Modulation}

\author{
\IEEEauthorblockN{Shiqiang Wang\IEEEauthorrefmark{1}\IEEEauthorrefmark{2}, Qingyang Song\IEEEauthorrefmark{1}, Lei Guo\IEEEauthorrefmark{1} and Abbas Jamalipour\IEEEauthorrefmark{3}}

\IEEEauthorblockA{\IEEEauthorrefmark{1}School of Information Science and Engineering, Northeastern University, Shenyang 110819, P. R. China}
\IEEEauthorblockA{\IEEEauthorrefmark{2}Department of Electrical and Electronic Engineering, Imperial College London, SW7 2AZ, United Kingdom}
\IEEEauthorblockA{\IEEEauthorrefmark{3}School of Electrical and Information Engineering, University of Sydney, NSW, 2006, Australia}
\IEEEauthorblockA{Email: shiqiang.wang11@imperial.ac.uk, songqingyang@ise.neu.edu.cn, guolei@ise.neu.edu.cn, a.jamalipour@ieee.org}

}

\maketitle

\begin{abstract}
\boldmath
The denoise-and-forward (DNF) method of physical-layer network coding (PNC) is a promising approach for wireless relaying networks. In this paper, we consider DNF-based PNC with $M$-ary quadrature amplitude modulation ($M$-QAM) and propose a mapping scheme that maps the superposed $M$-QAM signal to coded symbols. The mapping scheme supports both square and non-square $M$-QAM modulations, with various original constellation mappings (e.g. binary-coded or Gray-coded). Subsequently, we evaluate the symbol error rate and bit error rate (BER) of $M$-QAM modulated PNC that uses the proposed mapping scheme. Afterwards, as an application, a rate adaptation scheme for the DNF method of PNC is proposed. Simulation results show that the rate-adaptive PNC is advantageous in various scenarios.\footnote{˜\copyright˜ 2012 IEEE. Personal use of this material is permitted. Permission from IEEE must be obtained for all other uses, in any current or future media, including reprinting/republishing this material for advertising or promotional purposes, creating new collective works, for resale or redistribution 
to servers or lists, or reuse of any copyrighted component of this work in other works.}
\end{abstract}


\section{Introduction}
Physical-layer network coding (PNC) \cite{refFirstPatent, ref1, refanti-packets} has emerged as a new coding paradigm that can significantly improve the throughput performance of wireless relaying networks. The basic idea is to allow nodes to transmit simultaneously to the relay. After receiving a superposed signal, the relay performs a mapping operation to the superposed signal, and subsequently forwards the resulting signal to the destination nodes. The denoise-and-forward (DNF) method of PNC has been shown to outperform the amplify-and-forward (AF) method, because it avoids noise amplification \cite{refHanzoSurvey}.

\begin{figure}[!t]
\centering
\includegraphics[width=3.4in]{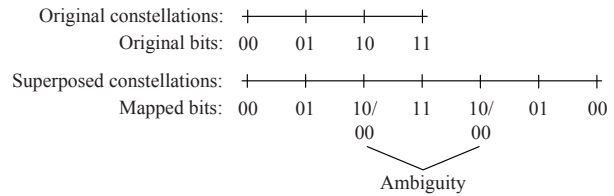}
\caption{Constellations and the corresponding bits with XOR mapping for 4-PAM modulation.}
\label{fig1}
\end{figure}

When using DNF with binary phase-shift keying (BPSK) or quadrature phase-shift keying (QPSK) modulations, the superposed signal can be mapped to the coded signal with the XOR operation \cite{ref1}. However, the XOR mapping method may not be suitable for higher level modulations. For instance, for the 4-ary pulse amplitude modulation (4-PAM), there exists ambiguity between the ``10'' and ``00'' bits when using XOR mapping, as shown in Fig. \ref{fig1}. This is also the case for $M$-ary quadrature amplitude modulation ($M$-QAM) with $M>4$, which is an extension to $M$-PAM. Therefore, we have to find alternative mapping methods for high-level modulations.

Because $M$-QAM is widely applied to rate-adaptive communication systems, it is significant that PNC supports general $M$-QAM modulations. A clustering-based mapping method that supports general $M$-QAM modulations for asynchronous DNF was proposed in \cite{ref7}, and a non-uniform constellation design for $M$-PAM to resolve the ambiguity problem was proposed in \cite{refHighOrderPAM}. However, the complexity of those methods are relatively high. In \cite{ref1}, a mapping scheme was proposed for synchronous DNF with $M$-QAM modulation. The scheme assumes that signal additions directly translate to symbol-wise modular $L$ (where $L=\sqrt{M}$) additions, which may not be suitable for Gray-coded symbols \cite{ref8} or non-square constellations.

Some other existing work focus on more theoretical aspects on PNC with high-level modulation and rate adaptation. Ref. \cite{ref10} proposed a precoding technique to improve the performance of PNC in rate-adaptive systems. A hierarchical modulation scheme to resolve the issue of asymmetric node positioning was proposed in \cite{ref11}. Also, the integration of lattice codes with PNC was discussed in \cite{reflattice}, which, however, is difficult to implement in practice \cite{refHighOrderPAM}.

In this paper, we focus on a practical solution to synchronous PNC with $M$-QAM modulation. We propose a simple but effective constellation mapping method, which supports both square and non-square constellations, as well as Gray-coded symbols (that are widely used in practical communication systems). Following that, we apply the proposed mapping scheme to a rate adaptive system, which selects the appropriate modulation level based on the channel status. Regarding the synchronization issues, note that we have a separate paper \cite{ref14} which focuses on phase-level synchronization for PNC.


The remainder of this paper is organized as follows. Section~II discusses the conditions of unique decodability and introduces the mapping scheme. Section~III evaluates the performance of PNC with $M$-QAM modulation using the proposed mapping scheme. The rate adaptation scheme is described and analyzed via simulations in Section~IV. Section~V draws conclusions.

\section{Constellation Mapping}
In this section, we discuss the conditions of unique decodability for $M$-PAM and $M$-QAM modulations and propose the corresponding mapping schemes based on constellation analysis.

\subsection{System Model}
\begin{figure}[!t]
\centering
\includegraphics[width=3.0in]{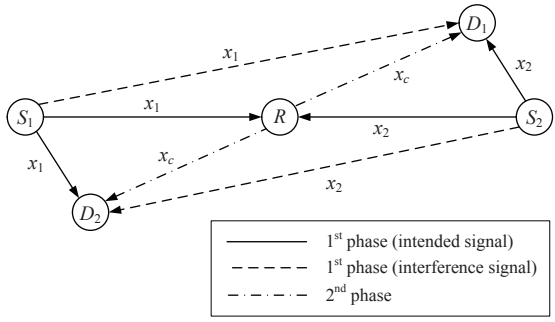}
\caption{Network topology.}
\label{fig2}
\end{figure}
We consider a network topology as shown in Fig. \ref{fig2}, in which $R$ denotes the relay, $S_1$ and $S_2$ denote the source nodes, $D_1$ and $D_2$ denote the corresponding destination nodes, $x_1$ and $x_2$ respectively denote the signals transmitted by the source nodes $S_1$ and $S_2$, and $x_c$ denotes the signal transmitted by the relay $R$.

The packet exchange process using PNC consists of two communication phases. In the first communication phase, the source nodes simultaneously transmit their packets to the relay. After the relay receives the superposed signal, it maps the superposed signal to a new signal, which carries a coded version of the packets. Then, in the second communication phase, the relay sends the coded packet to the corresponding destination nodes. From the coded packet, the destination can decode the packet it expects to receive, which is possible as long as the opportunistic listener $D_1$ (or $D_2$) has overheard the packet sent by the nearby source node $S_2$ (or $S_1$, correspondingly).

Let $s_1$ and $s_2$ respectively denote the two independent symbols that are carried by the signals $x_1$ and $x_2$, the mapping function $s_c=C(s_1, s_2)$ maps the superposed signal, which carries the symbols $s_1$ and $s_2$, to a coded symbol $s_c$, which is then broadcasted with the signal $x_c$. Note that, at the relay, only the superposed signal can be extracted. The relay is not aware of the individual values of $s_1$ and $s_2$. However, we use $s_1$ and $s_2$ as variables of the mapping function to simplify our subsequent analysis.

In this paper, we consider the case where the two signals that are superposed at the relay have synchronous phase and equal power. Hence, ignoring the channel attenuation and the noise, the superposed signal is $x_1+x_2$. Phase synchronization can be accomplished with the method described in \cite{ref14}. The equal power characteristic can be achieved by reducing the transmission power of the strong transmitter. As we will see in Section IV, this power reduction does not bring severe performance degradation for rate-adaptive systems, while it makes the mapping scheme design feasible.

\subsection{Mapping Scheme for $M$-PAM Modulation}
We first discuss the mapping scheme for $M$-PAM modulation. The mapping scheme has to be designed to ensure that the packets can be successfully decoded at the destinations. Based on the Exclusive Law \cite{ref7}, the necessary and sufficient condition for unique decodability at the destinations is as follows:
\begin{enumerate}
\item	For any $s_1' \neq s_1$, we have $C(s_1', s_2) \neq C(s_1, s_2)$.
\item	For any $s_2' \neq s_2$, we have $C(s_1, s_2') \neq C(s_1, s_2)$.
\end{enumerate}

Thus, we have the following proposition.
\begin{proposition}
\label{proposition1}
For $M$-PAM modulation, the necessary and sufficient condition for unique decodability is that any $M$ consecutive constellation points in the constellation of the superposed signal are mapped to different symbols.
\end{proposition}
\begin{IEEEproof}
The constellation of the superposed signal is a superposition of the constellations of the original signals. For two independent symbols $s_1$ and $s_2$, given $s_1$, the superposed signal $x_1+x_2$ has $M$ possible values. These $M$ possible values correspond to $M$ consecutive constellation points in the superposed constellation. The case is similar when $s_2$ is given. Hence, the statement that any $M$ consecutive constellation points are mapped to different symbols is equivalent to conditions 1) and 2) in the Exclusive Law. According to the Exclusive Law, the necessity and sufficiency are proved.
\end{IEEEproof}

Note that the constellation size of the superposed signal is $2M-1$. Let $s_{o,i}$\,$(i = 0, 1,\dots, M-1)$ denote the original symbol corresponding to the $i$\,th constellation point in the original constellation, and $s_{c,j}\,(j = 0, 1, \dots, 2(M-1))$ denote the coded symbol that the $j$\,th constellation point in the superposed constellation is mapped to. According to Proposition \ref{proposition1}, the value of the coded symbols can be evaluated by
\begin{equation}
\label{equa1}
s_{c,j}=s_{o,j\bmod M}\,.
\end{equation}

Eq. (\ref{equa1}) only shows one possible method of generating the coded symbols. In principle, any generation method that satisfies Proposition \ref{proposition1} can be applied. By generating the coded symbols according to (\ref{equa1}), we essentially map the superposed signal to symbols from the original symbol set.

Suppose the known symbol is $s_1=s_{o,i}$ and the received coded symbol is $s_c=s_{o,j \bmod M}$. For any $i$, $i' \in 0, 1, \dots, M - 1$, $i \neq i'$, we have $s_{o,i} \neq s_{o,i'}$. Hence, when $s_1$ is known, we can obtain the value of $i$. Similarly, at the destination, the value of $j \bmod M$ can be obtained from the received coded symbol $s_c$. The values of $i$ and $j$ can be regarded as scalars of the known signal $x_1$ and the superposed signal $x_1+x_2$, respectively. Therefore, the scalar of the signal $x_2$ is $j - i$. When the coded symbols are generated according to (\ref{equa1}), the superposed signal is mapped to symbols from the original symbol set by applying a $\bmod\text{ }M$ operation to its scalar. Hence, the symbol $s_2$, which the destination intends to receive, can be evaluated with a similar approach:
\begin{equation}
\label{equa2}
s_2=s_{o,(j-i)\bmod M}=s_{o,(j \bmod M-i)\bmod M}
\end{equation}

\begin{figure}[!t]
\centering
\includegraphics[width=3.0in]{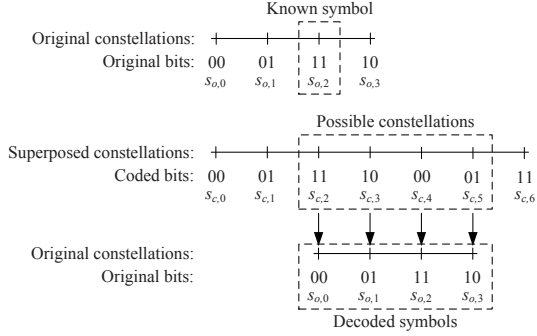}
\caption{Constellations, corresponding bits, and the decoding process of Gray-coded 4-PAM modulation with the proposed mapping method. Assume the known symbol is $s_{o,2}$, then the size of the possible superposed constellation is reduced to 4, and there is a one-to-one mapping between the coded symbol and the symbols that the destination expects to receive.}
\label{fig3}
\end{figure}

Eq. (\ref{equa2}) is applicable because the destination knows $j \bmod M$ and $i$. Fig. \ref{fig3} shows an example of the mapping and decoding process, when using Gray-coded 4-PAM modulation. As shown in Fig. \ref{fig3}, Proposition \ref{proposition1} ensures that, given a known symbol, there exists a one-to-one mapping between the coded bits and the expected bits. It can also be verified that the coded bits in Fig. \ref{fig3} are generated according to (\ref{equa1}) and the expected bits can be decoded by applying (\ref{equa2}).

\subsection{Mapping Scheme for $M$-QAM Modulation}
$M$-QAM modulation is a two-dimensional extension to $M$-PAM modulation. Hence, its conditions for unique decodability can be easily obtained from the Exclusive Law and Proposition \ref{proposition1}.
\begin{corollary}
\label{corollary1}
For $M$-QAM modulation with square constellations (i.e. the constellation size is $M=L \times L$, with $L$ being the number of constellation points on each of the in-phase and quadrature axes), the necessary and sufficient condition for unique decodability is that the constellation points in any $L$ by $L$ square in the constellation of the superposed signal are mapped to different symbols.
\end{corollary}
\begin{IEEEproof}
According to Proposition \ref{proposition1}, when only considering the in-phase component or the quadrature component of the superposed signal, the unique decodability condition is equivalent to: any $L$ consecutive constellation points (in terms of either the in-phase branch or the quadrature branch) are mapped to different symbols. It follows that, when jointly considering the in-phase and quadrature components, the unique decodability condition is equivalent to: the constellation points in any $L$ by $L$ square are mapped to different symbols.
\end{IEEEproof}

The generation method of coded symbols and the decoding process are similar with (\ref{equa1}) and (\ref{equa2}), except that two indexes (one for the in-phase component and one for the quadrature component) should be used. Let $s_{o,k,k'}$\,$(k, k' = 0, 1, \dots, L - 1)$ denote the original symbol corresponding to the $(k, k')$\,th constellation point in the original constellation, and $s_{c,l,l'}$\,$(l, l' = 0, 1, \dots, 2(L - 1))$ denote the coded symbol corresponding to the $(l, l')$th constellation point in the superposed constellation. A possible relationship between the coded and original symbols is
\begin{equation}
\label{equa3}
s_{c,l,l'}=s_{o, l \bmod L, l' \bmod L}\,.
\end{equation}

Suppose the known symbol is $s_1 = s_{o,k,k'}$ and the received coded symbol is $s_c=s_{o, l \bmod L, l' \bmod L}$, then the expected symbol $s_2$ can be decoded by
\begin{IEEEeqnarray}{ll}
\label{equa4}
s_2&{}={}s_{o, (l-k) \bmod L, (l'-k' ) \bmod L}\nonumber\\
&{}={}s_{o, (l \bmod L-k) \bmod L, (l' \bmod L-k') \bmod L}\,.
\end{IEEEeqnarray}
\begin{corollary}
\label{corollary2}
For $M$-QAM modulation with non-square constellations (but the vertical and horizontal spacings between constellation points still need to be equal or multiples of the minimum spacing), suppose the maximum number of constellation points seen on either the in-phase axis or the quadrature axis is $L'$, a sufficient condition for unique decodability is that the constellation points in any $L'$ by $L'$ square in the constellation of the superposed signal are mapped to different symbols.
\end{corollary}
\begin{IEEEproof}
For two symbols $s_1$ and $s_2$, given $s_1$ (or $s_2$), the possible constellation points of the superposed signal $x_1+x_2$ do not exceed an $L'$ by $L'$ square. Hence, when mapping the constellation points in any $L'$ by $L'$ square to different symbols, we have unique decodability.
\end{IEEEproof}

\begin{figure}[!t]
\centering
\includegraphics[width=3.0in]{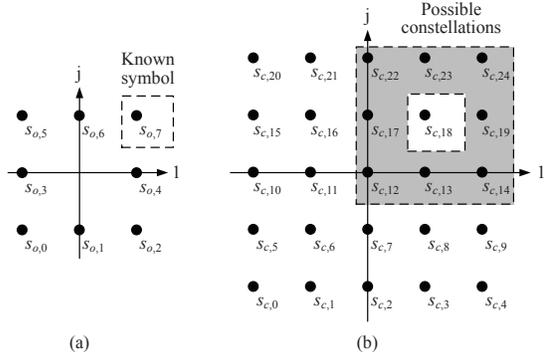}
\caption{Constellations of 8-QAM: (a) original signal, (b) superposed signal. Assume the known symbol is $s_{o,7}$, the possible superposed constellation points are those shown in the gray region. Because the superposed constellation points in any 3 by 3 square are mapped to different symbols, the destination can decode the expected symbol based on the coded symbol.}
\label{fig4}
\end{figure}
Note that Corollary \ref{corollary2} only provides a sufficient condition, because the number of the possible superposed constellation points is smaller than $L' \times L'$, when either $s_1$ or $s_2$ is given. This fact implies that there may exist mapping schemes that ensure unique decodability, in which some of the constellation points in an $L'$ by $L'$ square are mapped to the same symbol. However, Corollary \ref{corollary2} provides a simple way of designing the mapping scheme for non-square $M$-QAM modulation, such as 8-QAM (as shown in Fig. \ref{fig4}), 32-QAM etc.

For non-square $M$-QAM, the coded symbol set is generally larger than the original symbol set. For instance, with 8-QAM, the coded symbol set has $3 \times 3=9$ symbols; and with 32-QAM, the coded symbol set has $6 \times 6=36$ symbols. It follows that more bits are normally needed to transmit the coded symbol. However, considering that the theoretical throughput gain of PNC over conventional network coding (CNC) is 1.5 \cite{ref1}, PNC is theoretically still beneficial as long as the number of the additional bits is not larger than 50\,\% of the number of bits used to transmit the original symbol. It should also be noted that the additional bits need to be rearranged to match with the size of the original constellation.

The coded symbols for non-square $M$-QAM can be generated by extending the original constellation to a square constellation with $L' \times L'$ symbols. Then, the extended superposed constellation has $(2L' - 1) \times (2L' - 1)$ symbols. Based on the extended constellations, the generation of coded symbols and the decoding process for non-square $M$-QAM is the same with square $M$-QAM. The extended constellations are only used for calculation purposes, and the original constellations are still used for transmission. Consequently, some points in the extended constellations may not exist in actual transmissions.

\subsection{Complexity of Generating Coded Symbols}
For $M$-PAM or $M$-QAM, (\ref{equa1}) or (\ref{equa3}), respectively, must be evaluated for every superposed constellation point, to obtain all the relationships between the superposed constellation points and the coded symbols. The size of the superposed constellation is $2M - 1$ for $M$-PAM and $(2L - 1)^2 = 4M - 4M^{1/2} + 1$ for square $M$-QAM. For non-square $M$-QAM, note that the maximum possible value of $L'$ does not exceed $M$, hence the largest possible superposed constellation size is $(2M - 1)^2 = 4M^2 - 4M + 1$. Therefore, the complexity of the proposed coded symbol generation process for $M$-PAM or square $M$-QAM is $O(M)$; and for non-square $M$-QAM, it is $O(M^2)$. This complexity is lower than that of the clustering-based method \cite{ref7} (which is $O(M^4)$ due to its necessity of calculating and comparing the distances between constellation points) and the non-uniform constellation design method \cite{refHighOrderPAM} (which is $O(L!\times L!)$ because it needs to enumerate all the possible mappings).

\section{Performance with $M$-QAM Modulation}
In this section, we evaluate the symbol error rate (SER) and bit error rate (BER) performance of PNC with $M$-QAM modulation using the proposed mapping scheme, at different communication phases. Additive white Gaussian noise (AWGN) channel is assumed. The network topology under consideration is shown in Fig. \ref{fig2}. We assume that the receiver decodes the signals based on the minimum distance decision rule \cite{ref9}.

\subsection{Second Communication Phase}
We start our analysis with the second communication phase, where the symbols are sent with general $M$-QAM modulations. We summarize related theories regarding $M$-QAM in this subsection, to support our further discussions.

Let the distance between neighboring constellation points at the receiver be $2d$; as shown in \cite{ref8}, the SER is constrained by
\begin{equation}
\label{equa5}
p_s \leq 4Q\left(\sqrt{\frac{2d^2}{N_0}}\right)\,,
\end{equation}
where $Q(\cdot)$ is the tail probability of the standard normal distribution and $N_0$ denotes the power spectral density of noise.

For $M$-QAM with square constellations, according to \cite{ref8}, we can evaluate the exact SER by
\begin{equation}
\label{equa6}
p'_s=1-\left(1-\frac{2(L-1)}{L}Q\left(\sqrt{\frac{2d^2}{N_0}}\right)\right)^2\,,
\end{equation}
where $L=\sqrt{M}$.




Assume that the baseband signal is shaped with a raised cosine pulse with roll-off factor $\beta=1$, we can also write the SER with respect to the average signal-to-noise ratio (SNR) $\overline{\gamma}$ at the receiver \cite{ref8}:
\begin{equation}
\label{equa9}
p'_s=1-\left(1-\frac{2(L-1)}{L}Q\left(\sqrt{\frac{3\overline{\gamma}}{M-1}}\right)\right)^2\,.
\end{equation}

Because when an erroneous symbol is received, at least one bit is wrong and at most all the bits that the symbol carries are wrong, the BER $p_b$ is constrained by
\begin{equation}
\label{equa10}
\frac{p_s}{\log_2 M}\leq p_b \leq p_s\,.
\end{equation}

\subsection{Relay in the First Communication Phase}
In the first communication phase, the relay $R$ receives a superposed signal. For synchronous PNC with identical power, the superposed signal is still an $M$-QAM signal; and according to the discussions in Section II, adjacent superposed constellation points are mapped to different (coded) symbols. Hence, the SER is still constrained by (\ref{equa5}).

For square $M$-QAM, the number of superposed constellation points in either the in-phase branch or the quadrature branch is $2L - 1$. Ignoring the fact that there is repetition of coded symbols (i.e. different superposed constellation points can be mapped to the same symbol), the SER of the superposed signal can be obtained by substituting $2L - 1$ for $L$ in (\ref{equa9}), yielding
\begin{equation}
\label{equa11}
p'_{s\text{-}sup}\approx 1-\left(1-\frac{4(L-1)}{2L-1}Q\left(\sqrt{\frac{3\overline{\gamma}}{M-1}}\right)\right)^2\,.
\end{equation}

The BER constraint has the same form as (\ref{equa10}).

\subsection{Opportunistic Listeners in the First Communication Phase}
In the first communication phase, the destination $D_1$ (or $D_2$) overhears the symbols sent by its nearby source node $S_2$ (or, correspondingly, $S_1$). During this process, the signal sent by $S_1$ (or, correspondingly, $S_2$) can be regarded as interference. As a result, the distance between neighboring constellation points can be reduced. Let the neighboring distance in the constellation of the received interference signal be $2d'$, the maximum distance reduction is $2(L - 1)d'$, because the interference signal is in the same modulation as the intended signal. The maximum value is achieved when the two signals have a phase difference of $n\pi/2$ (where $n$ is an arbitrary integer) and the interference signal carries a symbol which lies on the outer boundary of the constellation.

According to the aforementioned discussions, we can obtain an upper bound of the SER:
\begin{IEEEeqnarray}{ll}
\label{equa12}
p_{s\text{-}opp}&{}\leq{}4Q\left(\sqrt{\frac{2(\max\{0,d-(L-1)d'\})^2}{N_0}}\right)\nonumber\\
&{}={}4Q\left(\max\{0,d-(L-1)d'\}\sqrt{\frac{2}{N_0}}\right)\nonumber\\
&{}={}4Q\!\left(\max\left\{0,\sqrt{\frac{2d^2}{N_0}}-(L-1)\sqrt{\frac{2d'^2}{N_0}}\right\}\!\right)\!\!.
\end{IEEEeqnarray}

For square $M$-QAM modulations, let
\begin{equation}
\label{equa13}
f(\alpha)=\max\left\{ 0, \sqrt{\frac{3\overline{\gamma}}{M-1}}-\alpha(L-1)\sqrt{\frac{3\overline{\gamma'}}{M-1}} \right\}\,,
\end{equation}
and
\begin{equation}
\label{equa14}
g(\alpha)=1-\left\{ 1 - \frac{2(L-1)}{L} Q\left( f(\alpha)\right) \right\}^2\,,
\end{equation}
where $\overline{\gamma'}$ is the average SNR of the interference signal at the receiver.

Then, the SER for square $M$-QAM is bounded by
\begin{equation}
\label{equa15}
g(0) \leq p'_{s\text{-}opp} \leq g(1)\,.
\end{equation}

The lower bound in (\ref{equa15}) has the same form with (\ref{equa9}). It corresponds to the case where no interference exists.

Considering that the symbols from the source nodes are equally probable, we can approximate the SER by taking an average value of the interference term in (\ref{equa13}), yielding
\begin{equation}
\label{equa16}
p'_{s\text{-}opp} \approx g(1/2)\,.
\end{equation}

At relatively high SNRs, it is more likely that only one bit error exists in an erroneous symbol. Hence, the BER for square $M$-QAM can be approximated by
\begin{equation}
\label{equa17}
p'_{b\text{-}opp} \approx \frac{g(1/2)}{\log_2 M}\,.
\end{equation}

\begin{figure}[!t]
\centering
\includegraphics[width=3.2in]{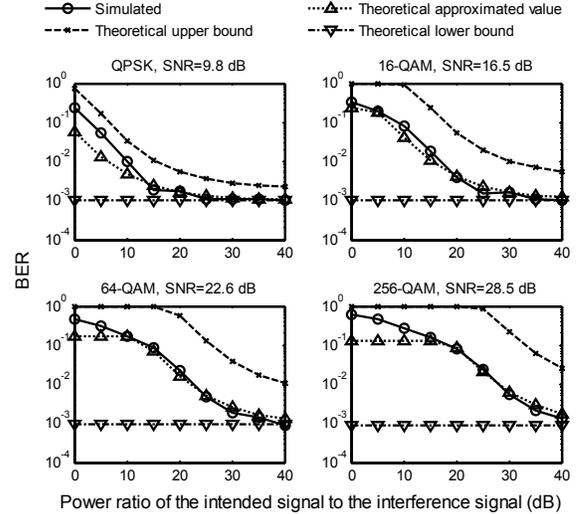}
\caption{BER at the opportunistic listeners in the first communication phase vs. power ratio of the intended signal to the interference signal, with various modulations.}
\label{fig6}
\end{figure}

The BER is further evaluated with simulations. We consider Gray-coded square $M$-QAM modulations with different values of $M$, and each source node sends 10,000 independent symbols. It is assumed that the phase difference between the intended signal and the interference signal is randomly distributed within $[0, 2\pi)$, because the two signals are independent with each other. The SNR of the intended signal (without interference) is set to a value ensuring that the lower bound of the BER is $10^{-3}$. We evaluate the BER under different power ratios of the intended signal to the interference signal. The results are shown in Fig. \ref{fig6}.

We can observe in Fig. \ref{fig6} that the simulated values approach the approximate values, particularly when the power ratio of the intended signal to the interference signal is high. Meanwhile, both of these values are constrained by the upper and lower bounds. When the SNR is low, the approximated values are constant, particularly when $M$ is large. This is because the approximated SER, which is evaluated from (\ref{equa16}), has reached its maximum value, causing the approximated BER also to reach its maximum value.

\section{Rate-Adaptive PNC}
By introducing $M$-QAM support for the DNF method of PNC, the data rate can be adjusted. In this section, we propose a simple rate adaptation scheme, in which the data is sent at the maximum possible data rate, under a specific channel condition and maximum BER requirement. The data rate is adjusted by using different modulations, including BPSK, QPSK, 16-QAM, 32-QAM (non-square), 64-QAM, 128-QAM (non-square), and 256-QAM. The rate adaptation scheme selects the highest level of modulation that satisfies the BER constraint. The upper bounds of the BER, as discussed in Section III, are used in this calculation.

\begin{figure}[!t]
\centering
\includegraphics[width=3.5in]{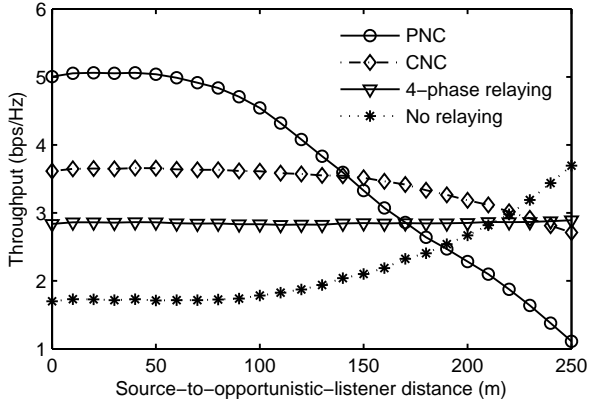}
\caption{Throughput vs. the distance between the source nodes and their corresponding opportunistic listeners, with rate-adaptive transmission.}
\label{fig7}
\end{figure}

The performance of the rate-adaptive PNC is evaluated with simulations. Its throughput is compared with other relaying methods, including CNC, conventional 4-phase relaying, and direct transmission without relaying. We also enable rate adaptation in the relaying methods that are used for comparison. The simulation settings are very similar with those in \cite{ref2}. The source nodes $S_1$ and $S_2$, and the relay $R$ are located on the same line, with the relay in the middle and the source nodes on opposite sides. The distance between each source node and the relay is uniformly distributed in $[125, 250]$\,m. The destination node $D_1$ (or $D_2$) is placed at a specific distance from $S_2$ (or $S_1$). We consider Rician flat-fading channels with Rician factor $K=5$\,dB and an average power gain of $1/r^4$, where $r$ is the distance between two nodes. In the simulations, the maximum transmission power is 10\,dBm, the noise power density is --174\,dBm/Hz, the noise figure is 6\,dB, and the receiver bandwidth is 1\,MHz. The maximum BER constraint is set to $10^{-3}$. Each simulation was run with 10,000 different random seeds to obtain the overall performance. We evaluate the performance for different distances between the source nodes and their corresponding opportunistic listeners. The results are shown in Fig. \ref{fig7}.

It can be observed in Fig. \ref{fig7} that, when the distance is smaller than 140\,m, PNC outperforms the other relaying methods. Specifically, when the distance is zero, which corresponds to the case where the source and destination nodes overlap, the throughput gain of PNC to CNC is 1.38, which is slightly lower than the theoretical value 1.5. The difference between the theoretical and simulated throughput gains is mainly due to the reduction of the transmission power of the strong transmitter when using PNC. However, as mentioned in Section~II, this power reduction makes it feasible to map the superposed signal to coded symbols.

The throughputs of both the PNC and CNC schemes decrease with the distance, because when the distance is large, the opportunistic listeners are less likely to overhear the symbols sent by the nearby source node. The throughput of PNC decreases faster, due to the presence of interference at the opportunistic listeners.

The throughput of transmitting without relaying increases with the distance, simply because when the distance is large, the source and destination nodes are closer to each other at certain random instances.

\section{Conclusion}
In this paper, we have been focusing on synchronous PNC supporting $M$-QAM. We have introduced a mapping scheme for superposed $M$-PAM and $M$-QAM signals, which has no restrictions on the original constellation mappings (it can be, for instance, either binary-coded or Gray-coded). Also, both square and non-square $M$-QAM modulations are considered. We have subsequently evaluated the performance of PNC with $M$-QAM modulation, both analytically and through simulations. The simulation results show that the theoretical upper and lower bounds, as well as the approximation values, are in accord with the simulated values. Following that, we have introduced a rate adaptation scheme for the DNF method of PNC. The rate adaptation scheme considers BPSK, QPSK, and $M$-QAM modulations. The appropriate modulation (and for $M$-QAM, the appropriate value of $M$) is selected according to the channel status and maximum BER constraint. Simulation results show that the rate-adaptive PNC outperforms other conventional rate-adaptive relaying schemes in various scenarios. A further analysis on theoretical aspects and optimization of rate adaptive DNF will be considered in our future work.

\section*{Acknowledgment}
We would like to thank Yang Huang for his suggestions and comments on this work.

This work was supported in part by the National Natural Science Foundation of China (61172051, 61071124), the Fok Ying Tung Education Foundation (121065), the Program for New Century Excellent Talents in University (08-0095, NCET-11-0075), the Specialized Research Fund for the Doctoral Program of Higher Education (20110042110023, 20110042120035), and the Fundamental Research Funds for the Central Universities (N100404008, N110204001, N110804003).

\bibliographystyle{IEEEtran}
\bibliography{MQAM_PNC_bib}


%


\end{document}